\documentstyle[pre,aps,preprint,epsfig]{revtex}

\tightenlines   
\begin{document}

\title{THE VISCOUS SLOWING DOWN OF SUPERCOOLED LIQUIDS
 AND THE GLASS TRANSITION: PHENOMENOLOGY, CONCEPTS, AND MODELS}
\author{G. TARJUS}\address{
Laboratoire  de Physique Th{\'e}orique des  Liquides, Universit{\'e} Pierre et
Marie Curie, 4 Place Jussieu, 75252 Paris cedex 05, France}
\author{D. KIVELSON}
\address{Department of Chemistry and Biochemistry,
University of California, Los Angeles, CA 90095, USA}
\maketitle
\begin{abstract}
The viscous slowing down of  supercooled  liquids that leads to  glass
formation   can  be considered  as a   classical,  and  is assuredly a
thoroughly studied, example of a "jamming process". In this review, we
stress the distinctive features characterizing the phenomenon. We also
discuss  the  main  theoretical approaches, with   an  emphasis on the
concepts    (free   volume,  dynamic    freezing    and  mode-coupling
approximations, configurational    entropy and     energy   landscape,
frustration) that could  be  useful in  other areas of   physics where
jamming  processes are encountered.
\end{abstract}
\vspace{0.5cm}

To appear in:
{\it Jamming and Rheology: Constrained Dynamics on Microscopic and Macroscopic 
Scales}\\
{\it S. Edwards, A. liu, and S. Nagel Eds.}

\section{INTRODUCTION}
When cooling a liquid,  usually under isobaric P=1atm  conditions, one
can often  bypass   crystallization, thereby obtaining   a supercooled
liquid  that    is metastable   relative  to  the  crystal.   When the
temperature is further  lowered, the viscosity of  the liquid, as well
as the relaxation times associated  with the primary ($\alpha$) relaxation
of   all kinds  of structural,   dielectric,  macro- and  micro-scopic
observables, increase rapidly, until a temperature is reached at which
the liquid can no longer flow and equilibrate in the time scale of the
experiment.   The  system effectively   appears  as a  rigid amorphous
material and is then called a glass. Glass formation thus results from
the    strong  viscous  slowing down   of   a  liquid  with decreasing
temperature\footnote{Although not as widely used,  there are other ways
of generating  glassy  structures,  such  as  vapor  deposition,  in situ
polymerization  or chemical reactions.\cite{R3}} ,   a slowing down that
can  be considered as  a classical example of   a ``jamming process''. A
characteristic  of this process    that is unanimously recognized,   a
unanimity rare in  this otherwise quite  open and controversial field,
is that it  is a dynamic effect.  The  so-called ``glass transition'' is
not a {\it bona fide} thermodynamic phase transition, but represents a
crossover  below  which a liquid   falls  out of   equilibrium on  the
experimental time scale.  The transition temperature, $T_g$ depends on
this time scale,  set either by  the observation  time (corresponding,
for  instance, to a  relaxation  time of  $10^2$ or  $10^3$  sec or  a
viscosity of $10^{13}$ Poise) and/or by the cooling rate (in a typical
differential  scanning calorimetry  measurement, $10$  K  per minute).
The dependence on cooling rate is,  however, weak, a difference of few
K in $T_g$  for an order-of-magnitude change of  the rate; this  is so
because on further  lowering   of the temperature  the   viscosity and
$\alpha$-relaxation times rapidly become enormous and out  of reach of any
experimental technique. 

In  the following, we shall focus  on the jamming process occurring in
the       supercooled     liquid     state.\cite{R1,R2}     Both   the
crystal\footnote{Note  that  for       several    liquids,  such    as
m-fluoroaniline  and  dibutylphtalate  at  atmospheric   pressure  and
atactic polymers, crystallization has  never been observed.}, which is
the stable phase  below the melting point  $T_m$ but can be ignored in
discussing the  glass transition, and  the glassy state itself will be
excluded from  the  present discussion.  By  appropriately eliminating
the  crystal  (experimentally  as  well as theoretically),  metastable
supercooled  liquids can  be  treated by   equilibrium thermodynamics,
statistical  mechanics, and conventional linear-response   formalisms.
Glasses  on   the other  hand,    although mechanically  stable,   are
out-of-equilibrium states;  especially   near    $T_g$, they   display
nonlinear responses and relaxations  known as aging or annealing,  and
their properties depend on  their history of preparation. \cite{R4,R5}
These phenomena will not be considered in this review.

\section{SALIENT   PHENOMENOLOGY }
The distinctive feature   of  glass-forming liquids is  the  dramatic,
continuous  increase of   viscosity  and  $\alpha$-relaxation  times  with
decreasing  temperature.  This sort of  jamming is observed in a large
variety   of  substances:  covalently   bonded  systems  like $SiO_2$,
hydrogen-bonded    liquids,  ionic   mixtures,   polymers,   colloidal
suspensions, molecular van der Waals liquids,  etc.  The emphasis will
be placed on those liquids (the vast majority) that  do not form 2- or
3-dimensional networks   of strong bonds because   they  show the most
striking behavior when  passing from the high-temperature liquid phase
to the deeply supercooled and very viscous regime.

\subsection{Strong, super-Arrhenius  T-dependence of
viscosity  and   $\alpha$-relaxation times    }  The viscosity   $\eta$  and
$\alpha$-relaxation  times can change  by $15$ orders   of magnitude for a
mere decrease of  temperature  by  a factor two\footnote{The    (shear)
viscosity   $\eta$   can be   related   to  a   time characteristic  of
$\alpha$-relaxation, the average  shear stress relaxation time  $\tau_s$, by
$\eta=G_\infty  \tau_s$,   where  $G_\infty$ is the infinite-frequency shear
modulus; $G_\infty$ is    typically   of   the order   of
$10^{10}-10^{11}$ erg.cm$^{-3}$, so that  a viscosity of $10^{13}$
Poise roughly
corresponds to a time  of  $10^2$  or  $10^3$ sec.}. Such  a  dramatic
variation is conveniently represented on a logarithmic plot of $\eta$ or
$\tau_\alpha$ versus $1/T,$  i.e., an Arrhenius plot:  see  Fig.   1a.  A
system like $GeO_2$, an   example  of  a network-forming   system,  is
characterized  by   an  almost linear  variation,  which  indicates an
Arrhenius temperature dependence.  For all other liquids on the figure
there     is   a  marked  upward     curvature,   which   represents a
faster-than-Arrhenius, or  super-Arrhenius, temperature dependence and
is often described by an empirical Vogel-Fulcher-Tammann (VFT) formula
(also called Williams-Landel-Ferry formula  in the context of polymers
studies\cite{R6}),
\begin{equation}
\label{eq:1}
\tau_\alpha =\tau_0\exp\left(D\frac{T}{T-T_0}\right),
\end{equation}
where   $\tau_0$, $D$ and  $T_0<T_g$  are adjustable  parameters. On the
basis of  such Arrhenius plots, with the  temperature scaled by $T_g$,
Angell   proposed  the  now   standard classification of glass-forming
liquids  into   {\it  strong}    (Arrhenius-like) and  {\it   fragile}
(super-Arrhenius)  systems;\cite{R7} in Eq.~(\ref{eq:1}), the smaller
the value of $ D$, the more fragile the liquid.  There are, of course,
alternative fitting  formulas that have been used,  some  which do not
imply a singularity at a nonzero temperature as does the expression in
Eq.~(\ref{eq:1}).\cite{R8}

A  different   way of representing  the    phenomenon is to   plot the
effective activation free energy for $\alpha$-relaxation, $E(T)$, obtained
from

\begin{equation}
\label{eq:2}
\tau_\alpha =\tau_{\alpha,\infty}\exp\left(\frac{E(T)}{k_BT}\right),
\end{equation}
where $k_B$ is  the Boltzmann constant and  $\tau_{\alpha,\infty}$ is a high-$T$
relaxation time, or from a similar equation for the viscosity. This is
illustrated in Fig. 1b, where the temperature has been scaled for each
liquid  to a  temperature  T* above  which  the dependence is  roughly
Arrhenius-like.  Although  the  determination of $T^*$ is   subject to
some  uncertainty,\cite{R8,R9} the  procedure emphasizes the crossover
from Arrhenius-like to super-Arrhenius behavior that is typical of and
quite distinct  in most supercooled  liquids.  The appreciable size of
the effective activation free  energies $E(T)$, namely, $40 k_BT_g$ at
the glass    transition,    is  indicative   of thermally    activated
dynamics.  Such a large   effective activation free energy  for weakly
bonded  fragile   molecular  liquids  such as  orthoterphenyl    is an
intriguing feature of the phenomenology.  Another peculiar property of
$E(T)$ for fragile systems is  that it increases significantly between
$T^*$ and $T_g$ (a factor $3$, i.e., a factor of  $5$ or $6$ in units
of  the  thermal energy $k_BT$,  for weakly   bonded fragile liquids).
Such a variation is not  commonly encountered.   For instance, in  the
field of critical phenomena, the slowing  down of dynamics that occurs
when approaching  the critical point  is  usually characterized  by  a
power law  growth  of  the  relaxation time;  in   terms of  effective
activation  free energy, this  corresponds to a logarithmic growth and
it  is  slower  than  the  variation described   by  the VFT  formula,
Eq.~(\ref{eq:1}).  Unusually strong  slowing  down, with exponentially
growing times similar to Eq.~(\ref{eq:1}), is found in some disordered
systems  like the  random   field Ising   model and  it  is known   as
``activated dynamic    scaling''.\cite{R10}

\subsection{Nonexponential         relaxations}
In an ``ordinary'' liquid  above the melting point, relaxation functions
are usually  well described,  after some transient  time, by  a simple
exponential decay.  Deviations are  observed,   but they are   neither
systematic   nor  very   marked.   The    situation changes at   lower
temperatures, and the $\alpha$-relaxation is no longer characterized by an
exponential   decay.   A   better representation  is    provided  by a
``stretched exponential'' (or Kohlrausch-Williams-Watts function),

\begin{equation}
\label{eq:3}
f_\alpha(t)\propto \exp\left[-\left(\frac{t}{\tau_\alpha}\right)^\beta \right],
\end{equation}
where $\beta$  is  the stretching parameter;   the smaller $\beta$  the more
``stretched''  the relaxation.  Although  not unambiguously established,
the degree  of departure   from exponential behavior,   or stretching,
appears  to  increase   (i.e., $\beta$  decreases)   with  decreasing
temperature.  

Alternatively, in frequency space, the spectrum  of the imaginary part
of the susceptibility, which is characterized by a  peak at a frequency
$\omega_\alpha \propto 1/ \tau_\alpha $, tends to be broader (when  plotted as a function
of $log(\omega)$  ) than the  simple Lorentzian or  Debye spectrum that is
just the  Fourier transform of  the  time-derivative of an exponential
relaxation   function  (see  Fig.  2).   Fitting  formulas  related to
Eq.~(\ref{eq:3}),  formulas  like   the  Cole-Davidson function    for
frequency-dependent susceptibilities, $(1-i(\omega/\omega_\alpha ))^{-\beta'}$, are used
to fit  the spectroscopic data,  but similar trends  are observed: the
$\alpha$ peak, as  observed  for instance   in the imaginary  part of  the
dielectric susceptibility as a  function of $log(\omega)$, broadens as the
temperature  is  lowered  towards $T_g$,\cite{R11}   which   indicates
increasing   departure from  Debye/exponential  behavior.   Except for
network-forming  systems,  the stretching of  the   $\alpha$ relaxation is
significant ($\beta$ is  typically  between $0.3$  and $0.6$ for  fragile
liquids at $T_g$).  However, and  this point may  not have been  given
enough attention, the stretching, or broadening in frequency space, is
relatively small  when compared to the  extremely rapid variation with
temperature of  the  $\alpha$-relaxation  time  itself.   This  is  to  be
contrasted for instance  with the activated critical slowing discussed
above. There, the power law growth  of the activation free energy when
the  temperature is decreased toward the  critical  point comes with a
more striking stretching of  the relaxation function  that occurs on a
logarithmic scale:  in this case,  in place of a stretched exponential
behavior as in Eq.~(\ref{eq:3}), $\ln(f(t))$  goes  as some power of  
$(1/\ln(t))$.\cite{R12}
\subsection{No  marked  changes  in  structural  quantities}

It is tempting to associate the huge increase in $\alpha$-relaxation times
and  viscosity with  the growth   of  a {\it structural}   correlation
length.  However,   no such growth   has been   detected so   far   in
supercooled liquids.   Quite    to  the contrary,   the   variation of
structure in  liquids  and glasses, as measured  in  neutron and X-ray
diffraction experiments,  appear rather  bland\cite{R13,R14} (see Fig.
3).  The ordinary,  high-temperature liquid has only short-range order
whose signature in the static structure  factor $S(Q)$ is a broad peak
(or a split peak for some molecular systems as illustrated in Fig.  3)
at a wave vector  $Q$ that roughly corresponds in  real space to  some
typical   mean   distance  between  neighboring   molecules.  As   the
temperature is lowered and   the supercooled regime is entered,  there
are small,  continuous variations of  the structure factor that mostly
reflect the change in density (typically, a $5\%$ change between $T_m$
and $T_g$) and, possibly,  some adjustments in the local  arrangements
of the molecules.  There  is  no sign,   however, of  a  significantly
growing correlation length, nor of the appearance of a super-molecular
 length. 

In  network-forming  and
$H$-bonded systems, an  additional ``pre-peak'' is sometimes detectable at
wave  vectors somewhat  lower than  that of the  main peak,  but it is
attributed  to specific effects  induced  by the strongly  directional
intermolecular  bonds and not to  a length  scale that would correlate
with  the   viscous   slowing  down.\cite{R15}

In
contrast to this lack of structural  signature for the existence of an
increasing  super-molecular   correlation     length   with decreasing
temperature, there  is significant evidence,  as discussed below, that
corresponding ``dynamical''  correlation length do  exist.

\subsection{Rapid entropy decrease and Kauzmann paradox }

The absence of marked changes  in the structure, {\it  at least at the
level  of two-particle density correlations}, or  of a strong increase
in  any directly measured static  susceptibility is a puzzling feature
of  the jamming  process  associated  with  glass formation. The  only
static quantity that  shows behavior  that might  be relevant is   the
entropy.  Below the  melting point, $T_m$,  the heat capacity $C_p(T)$
of a supercooled  liquid is  larger   than that of the   corresponding
crystal. (At $T_g$, the $C_p$ of  the liquid drops  to a value that is
characteristic of the glass and is close to the  $C_p$ of the crystal,
but this is  a  consequence of the   system no longer being   properly
equilibrated.) As a result of this ``excess'' heat capacity, the entropy
difference  between  the liquid  and     the crystal decreases    with
temperature, typically by a factor of $3$ between  $T_m$ and $T_g$ for
fragile liquids. The effect is illustrated in Fig.  4 and leads to the
famous Kauzmann   paradox:\cite{R16}  if  the  entropy   difference is
extrapolated  to  temperatures below  $T_g$,  its  extrapolated  value
vanishes  at   some nonzero temperature  $T_K$,   which results in the
unpleasant  feature that the  entropy of the  liquid  becomes equal to
that  of the  crystal (even  more  unpleasant: if the extrapolation is
carried to still lower temperatures, the entropy of the liquid becomes
negative, which violates the third law of thermodynamics). The paradox
is that this {\it extrapolated }entropy crisis is avoided for a purely
dynamic  reason, the intervention  of the glass transition: what would
occur if  one were  able to keep  the supercooled  liquid equilibrated
down to temperatures  below $T_g$?  There  are certainly  many ways to
answer the  question. The paradox  could be resolved  by the existence
between $T_g$ and $T_K$ of an  intrinsic limit of metastability of the
liquid\cite{R16} or of  a  second-order phase transition\footnote{Note
that a  low-$T$  first-order transition  does not resolve  the paradox
because it can be  supercooled.}  (a speculation that gains additional
credibility with the  observation  that the VFT  temperature  $T_0$ at
which the  extrapolated  viscosity and  $\alpha$-relaxation  times diverge
(see     Eq.~(\ref{eq:1}))      is     often    found    close      to
$T_K$\cite{R17,R18}).  Even  more  simply,  one might   find  that the
extrapolation breaks down above  $T_K$ and that the entropy-difference
curve levels off and  goes smoothly to zero  at zero $K$, in  much the
same  way as it  does in the Debye  theory of crystals.   These are of
course all speculations, but it remains that the rapid decrease of the
entropy of the   supercooled liquid relative   to that of  the crystal
represents an intriguing    aspect  of the phenomenology   of  fragile
glass-formers.

\subsection{Two-step relaxation and secondary processes}
As we stressed before, the salient features related to glass formation
concern the long-time (low-frequency) primary or  $\alpha$ relaxation.  As
the $\alpha$-relaxation time increases with decreasing temperature, so too
does the window between this  time and typical microscopic, picosecond
or  sub-picosecond times.  When   the  relaxation function  is plotted
against the logarithm of the time, one then observes what is sometimes
called a ``two-step  relaxation''.  An illustration is given  in Fig.  5
by the dynamic  structure  factor  of the fragile   ionic glass-former
$Ca_{0.4}K_{0.6}       (NO_3)_{1.4}$    obtained     by        neutron
techniques.\cite{R19} At  high temperature, the relaxation function is
essentially a  one-step process.  However,  as the liquid becomes more
viscous, the relaxation proceeds in two steps  separated by a plateau.
Although the terminology is  far from being universally  accepted, the
approach to the plateau from the short-time side  is often referred to
as $\beta$ or fast-$\beta$ relaxation.  If one is  to fit the long-time part
by a stretched exponential  (Eq.~(\ref{eq:3})), there is a large range
of ``mesoscopic'' times that is not adequately described and that widens
as the temperature is lowered.  Power  law functions of time are often
used to reproduce the relaxation function in this mesoscopic range.

This two-step relaxation feature is common to all fragile liquids.  In
addition,  there  may  also  appear   additional secondary  processes,
detected  first by  Johari  and Goldstein in  dielectric spectroscopy.
\cite{R20}  Such  secondary processes,  whose   presence and  strength
strongly   vary from  one   liquid  to  another,  have  characteristic
frequencies that  are  intermediate  between  those of the   $\alpha$  and
fast-$\beta$   relaxations.   They  are  denoted Johari-Goldstein-$\beta$,
slow-$\beta$,  or simply  $\beta$    processes. \cite{R2,R21}  To   make the
description more complete,  one  should  also mention  the   so-called
``boson   peak'' that may     be  present on  the  high-frequency   side
($\sim  10^2-10^3Ghz$) of light  and   neutron scattering (or   absorption)
spectra.\cite{R2,R22} Here we  do not discuss either the slow-$\beta$
  processes or the boson
peak.   

\section{ A SELECTION OF QUESTIONS }
After this brief review of the salient aspects of the phenomenology of
supercooled liquids as they  get glassy, we discuss  in more detail  a
number  of  questions,  whose    answers  give  justification or   put
constraints on the  theoretical picture one  can build to  explain the
viscous slowing down. 
\subsection{How  universal   is the  behavior   of  glass-forming
liquids ?}

Universality  is a  key concept  in  physics and it  has  proven to be
central  in the   field of  critical phenomena.   By  the standards of
critical phenomena  studies, the  observed  behavior  of glass-forming
liquids is not universal, the main reason being that no singularity is
detected  experimentally  (or approached   asymptotically  close),  as
stressed above.  However, if one is willing to take  a broader view of
the  notion of universality,  one  can find considerable generality or
``universality''  in   the   properties, those  mostly   associated with
long-time and  low-frequency phenomena, that characterize the approach
to   the   glass  transition.    For  instance,  the   super-Arrhenius
$T$-dependence of   the  viscosity and  $\alpha$-relaxation times  and the
nonexponential character of the  relaxation function are  observed for
virtually all glass-formers, be they polymeric, $H$-bonded, ionic, van
der Waals,    etc.,  with  the exception  of    a minority   of strong
network-forming systems; and, for a given liquid, these properties are
found  by  a   large  variety  of experimental   techniques,   such as
dielectric  relaxation,  light and neutron  scattering, NMR, viscosity
measurements,  specific   heat    spectroscopy,  volume  and  enthalpy
relaxation, optical probe methods.

The presence of an underlying "universality" is  supported by the fact
that experimental data  covering a wide range   of temperatures and  a
great diversity of substances can be collapsed onto master curves with
only a  small number  of species-dependent adjustable   parameters.  A
good   example   is  provided    by    the  scaling   plot   of    the
frequency-dependent dielectric susceptibility  proposed  by  Nagel and
coworkers.\cite{R11} As  shown in Fig.  6, the  data taken over a $13$
decade range of  frequencies for many different  liquids can be placed
with good accuracy  onto a single curve after  scaling with only three
parameters associated     with  the $\alpha$-peak     position,  width and
intensity.  The  master-curve  for the  temperature  dependence of the
effective activation free energy for viscosity and $\alpha$ relaxation put
forward  by Kivelson {\it et al.}  \cite{R8} is another example. It is
nonetheless fair  to  say that  the fits  resulting from these various
scaling procedures are far from perfect, which  leaves room for debate
and conflicting interpretations. One can also ask the question whether
the universality holds only  up to the implied high-frequency  cut-off
of   the  susceptibility  scaling   curve  of   Nagel   {\it  et  al.}
(i.e., whether one should  be  focusing on slow behavior) or
whether it extends higher in light of the  fact that similarities have
also    been  observed     in    the  high-frequency susceptibilities.
\cite{R23}
\subsection{  Is   the $\alpha$-relaxation homogeneous or  heterogeneous?}
The  observation    stressed above     that the   $\alpha$-relaxation   is
nonexponential in the supercooled liquid  range and its representation
by a stretched   exponential  as in Eq.~(\ref{eq:2})  can  be formally
interpreted  in terms  of  a superposition  of  exponentially decaying
functions with a distribution of relaxation times;  but, this {\it per
se} does  not guarantee that  the dynamics be  ``heterogeneous'', in the
sense that relaxation of the molecules differs from one environment to
another  with  the   environment life time   being    longer than  the
relaxation time.   An alternative explanation  can be offered within a
``homogeneous'' picture in which  relaxation of the molecules everywhere
in the liquid is intrinsically nonexponential.

In  recent  years,   there  has     been   mounting evidence      that
heterogeneities,    sufficiently long-lived to   be   relevant to  the
$\alpha$-relaxation   and  to be   at  least  partly  responsible  for its
nonexponential feature, do exist in supercooled liquids.\cite{R24} The
heterogeneous character of the slow  dynamics has been demonstrated in
several experiments:  multi-dimensional NMR, \cite{R25} photobleaching
probe rotation measurements,  \cite{R26}  nonresonant  dielectric hole
burning.  \cite{R27} These  techniques   involve  the selection  of  a
sub-ensemble  of molecules  in the  sample  that is characterized by a
fairly  narrow distribution  of   relaxation times (and  in general  a
relaxation  slower  than average)  and the further  monitoring  of the
gradual return to the  equilibrium situation.  Additional  evidence of
the spatially   heterogeneous   nature  of the   dynamics   in fragile
supercooled  liquids is provided by   the  so-called breakdown of  the
Stokes-Einstein relation  between the translational diffusion constant
and the viscosity, and the concomitant ``decoupling'' between rotational
and   translational    time  scales:\cite{R28,R29} see  Fig.7.

The  size of the  heterogeneities  is not  directly observable in  the
above mentioned experiments, but  various estimates, obtained, e.  g.,
from optical studies of the rotational relaxation of probes of varying
size,  \cite{R30} NMR   measurements, \cite{R31} the   study of excess
light scattering,  \cite{R32}  and  the  influence of  a  well-defined
$3$-dimensional confinement \cite{R33} lead   to a typical length   of
several nanometers in different fragile liquids near $T_g$. One should
recall that these signatures  are all dynamical  and that no signature
at such a length scale has been detected so far in small-angle neutron
and X-ray diffraction data.    If the heterogeneous character  of  the
$\alpha$-relaxation  appear  reasonably  well  established,  at  least for
deeply  supercooled  fragile  liquids,  several points  concerning the
lifetime, the size, and the  nature of the heterogeneities need  still
be settled.

\subsection{Is  density or temperature the dominant control variable?}
  The phenomenon of viscous
slowing down and glass  formation as it  is  studied most of the  time
(and  described in the  preceding sections) takes place under isobaric
$P=1atm$ conditions. As a consequence, when  the temperature is lowered,
there is also an increase of the density of the liquid.  This increase
is     small   (a typical  variation   of     $5\%$  between   $T_m$
and $T_g$),   but it could still  have  a
major influence   on the dynamics.    Actually,  there are theoretical
models of  jamming, such as  those  based on free  volume concepts and
hard   sphere  systems, that  attribute   the spectacular  increase of
viscosity and $\alpha$-relaxation
times of    fragile glass-formers (almost)    entirely to  the density
changes. It  is  thus important to  evaluate the  role of density  and
temperature in driving  the jamming process  that  leads to the  glass
transition {\it  at 1 atm}. 

Basic  models and theories are  usually formulated  in terms of either
density  or temperature as     control variable, but experiments   are
carried out with  pressure   and   temperature as external     control
variables.   The  data must be   converted,  when enough  experimental
results are available, in order to analyze the influence of density at
constant temperature and that of  temperature at constant density, for
a  range of density  and  temperature  that  is characteristic of  the
phenomenon  {\it at  1 atm}.   Extant  analyses\cite{R34} are far from
exhaustive.  However, as  illustrated  in Fig.  8, the  characteristic
super-Arrhenius   T-dependence     of  the   viscosity,   $\eta$,    and
$\alpha$-relaxation  times,  $\tau_\alpha$, appears   predominantly due  to  the
variation of temperature and not to  that of density.  This conclusion
is confirmed  by a comparative  study of the  contributions induced by
density variations  (at    constant temperature)  and   by  temperature
variations (at constant  density) to the rate of  change  of $\eta$ and
$\tau_\alpha$   at constant (low) pressure  in  the viscous liquid regime of
several molecular and polymeric glass-formers. \cite{R34}

How general   is  the   above  result? Temperature
appears  to  be the    dominant   variable controlling  the    viscous
super-Arrhenius slowing down  of supercooled liquids at  low pressure,
but  this may not be  the case at  much  higher pressure (although not
much data  are presently available  to confirm this  point), and it is
most  likely     not true  for    describing the  concentration-driven
congestion  of dynamics in colloidal suspensions.  In the absence of a
``super-universal'' picture of   the jamming  associated  with
glass formation, we shall  restrict  ourselves, as we have  implicitly
done above, to the consideration of supercooled  liquids at 1 atm.

\subsection{What  are  the relevant characteristic temperatures?}

There  is   no  unbiased way  of    presenting  the  phenomenology  of
glass-forming liquids. Choices must be made  about the emphasis put on
the  different aspects, about  the  best graphic representations,  and
about the way in which one analyzes  experimental data.  To make sense
out of the wealth of  observations and measurements,  it is natural to
look for characteristic temperatures about which to organize and scale
the   data.  However, since no singularity   is directly detected, the
selection  of  one   or several relevant    temperatures  is far  from
straightforward.   The temperatures   that  can be  easily  determined
experimentally are the boiling point, $T_b$, the melting point, $T_m$,
and the  glass transition  temperature(s), $T_g$.  Unfortunately,  the
former  two are  generally   considered as irrelevant  to the  jamming
phenomenon, and    the   latter has  an   operational  rather   than a
fundamental  nature (see the introduction). 

Several other candidates have  been suggested, that  can be split into
two groups.  First, there are ``extrapolation temperatures'' {\it below}
$T_g$, i.e.,  temperatures dynamically  inaccessible to supercooled
liquids, at which  extrapolated behavior diverges or becomes singular.
This  is the case  of the VFT temperature  $T_0$ (see Eq.   1) and the
Kauzmann  temperature $T_K$  (see section II-4  and  Fig.  4).  In the
second group are ``crossover temperatures'' {\it above} $T_g$ at which a
new  phenomenon seems     to  appear (decoupling    of  rotations  and
translations, emergence    of  a secondary    $\beta$  process,  etc.), a
crossover of behavior or a change of 
$\alpha$-relaxation  mechanism seems  to take place   (passage from
Arrhenius  to super-Arrhenius  $T$-dependence, arrest  of the relaxation
mechanisms described by the  mode-coupling theory, putative  emergence
of activated barrier crossing  processes,  etc.).  A variety  of  such
temperatures for the fragile  glass-former $OTP$ are shown  in Fig.   9:
the   location  of   the   different characteristic    temperatures is
illustrated  on an Arrhenius plot  of the viscosity. It is interesting
to note that most putative crossover behaviors occur  in the region of
strong curvature where $\eta \sim  1-10^2$ Poise or $\tau_\alpha \sim 10^{-10}-10^{-8}$
sec,  while,   on  the  other   hand,  the   temperatures obtained  by
extrapolation of data to low T lie fairly close to each other, some 40
K below $T_g$.

 \subsection{What  can be learned from computer simulations?}

Computer  simulation studies, \cite{R39}  in particular those based on
Molecular Dynamics  algorithms,   have proven extremely   valuable  in
investigating the structure and  dynamics of simple, ordinary liquids.
Their  contribution  to the  understanding of  the glass transition is
unfortunately limited, the  main reason being  the restricted range of
lengths  and times  that    are   accessible to Molecular     Dynamics
simulations:  typical   simulations on    atomistic   models  consider
$10^3-10^4$  atoms and can follow  relaxations for less than $10^{-8}$
sec  (when expressing the elementary  time step in terms of parameters
characteristic of simple liquids).   As a result, the viscous,  deeply
supercooled  regime   of  real  glass-forming  liquids,   where strong
super-Arrhenius   behavior,   heterogeneous   dynamics,    and   other
significant features associated  with the jamming process develop,  is
out of reach, as  is the laboratory glass transition  that occurs on a
time  scale of $10^2$   or $10^3$ sec.  Simple   liquid models do form
``glasses'' on  the observation, i.e.,  simulation, time with  many of
the attributes of the laboratory   glass transition: abrupt change  in
the thermodynamic coefficients, dependence on  the cooling rate, aging
effects,   etc\footnote{ However,    these   models   are   effectively
high-temperature  structures,  since     they correspond   to   liquid
configurations that are kinetically arrested at a temperature at which
the primary relaxation time is of the order  only of nanoseconds. Even
with systems   specially designed to  avoid  crystallization,  such as
binary Lennard-Jones  mixtures, the cooling  rates  to prepare glasses
($10^8$ to $10^9$   K/sec) are orders  of  magnitude higher than  those
commonly used in experiments.}.  However, these supercooled simulation
liquids  are  not   truly    fragile   (the  $T$-dependence  of    the
$\alpha$-relaxation   time shows  only    small departure  from  Arrhenius
behavior) and   are not deeply supercooled so    that one{\'{}}s ability to
extract   insights into  the  deeply   supercooled fragile liquids  is
questionable.

Computer   simulations  can  be  useful   in  studying the  moderately
supercooled liquid region, where one can  observe the onset of viscous
slowing down (see for  instance Fig.  9). The  major interest  of such
studies is    that  static  and   dynamic   quantities that   are  not
experimentally accessible can  be   investigated, such  as  multi-body
(beyond  two-particle) correlations  involving  a variety of variables
and microscopic mechanisms  for  transport and  relaxation.  They also
allow for  testing  in  detail  theoretical  predictions  made in  the
relevant   window of times   and    lengths (e.   g.,   those of   the
mode-coupling theory) and   for analyzing properties   associated with
configurational or phase   space (see below).  

In addition to  the  much studied   one- or two-component  systems  of
spheres  with spherically  symmetric interaction potentials,\cite{R39}
the  models investigated in  computer simulations  can be divided into
two main  groups: on one hand, more   realistic microscopic models for
molecular   glass-formers  that attempt   to describe species-specific
effects;\cite{R40}   on  the  other   hand,  more  schematic  systems,
coarse-grained representations, \cite{R41} lower-dimensional systems
\cite{R42} or toy-models,\cite{R43}
that  bear less detailed resemblance with  real glass-formers, but can
be studied on much longer time scales and with bigger system sizes.

\section{THEORETICAL APPROACHES.}
There is  a  large number  of  theories,  models, or simply  empirical
formulae that  attempt to  reproduce  pieces  of the phenomenology  of
supercooled liquids.  There  are   fewer  approaches,  however,   that
address the question of why  and how the  viscous slowing down leading
to the glass transition, with its salient characteristics described in
the preceding sections, occurs  in liquids as they  are cooled. In the
following,  we shall briefly review  the  main theoretical approaches,
with an emphasis on the concepts and  methods that may prove useful in
other areas of physics  where  some sort  of  jamming process is  also
encountered\footnote{Models addressing  more specific questions,  such
as the  ``coupling model'' \cite{R44}  or the ``continuous  time random
walk'', approach\cite{R45}  are   discussed in the   reviews  cited in
\cite{R1}.}.   More  specifically, we  shall  discuss phenomenological
models  based  on free    volume   and configurational   entropy,  the
description of  a purely dynamic   arrest resulting from mode-coupling
approximations, ideas relying on the  consideration of the topographic
properties of   the  configurational  space   (energy and  free-energy
landscapes) or on the analogy with  generalized spin glass models, and
approaches centered on the concept of frustration.

\subsection{Free volume } 
Free-volume models rest on the assumption  that molecular transport in
viscous fluids occurs only when voids having a  volume large enough to
accommodate   a molecule form   by  the  redistribution of  some ``free
volume'', where  this latter is  loosely defined as some surplus volume
that is not  taken up by the  molecules.  In the standard presentation
by Cohen and Turnbull,\cite{R46} a molecule in a   dense fluid is mostly
confined to a cage  formed by its  nearest neighbors.  The  local free
volume,  $v_f$,  is roughly that part  of  a cage space
which  exceeds that taken  by a molecule.  It  is assumed that between
two events contributing to  molecular transport, a reshuffling of free
volume among  the cages occurs  at  no cost of  energy.  Assuming also
that the  local  free  volumes  are  statistically  uncorrelated,  one
derives a   probability distribution, $P(v_f)$, which is exponential,
\begin{equation}
\label{eq:4}
P(v_f)\propto \exp\left(-\gamma \frac{v_f}{\overline{v_f}}\right),
\end{equation}
where $\overline{v_f}$ is the average free volume per  molecule and $\gamma$ is
a  constant  of  order  $1$. Since    the limiting mechanism   for the
diffusion of a molecule is  the occurrence of a void,  i.e., a  local
free  volume  $v_f$  larger  than some   critical value, $v_0$, that is
approximately equal to  the molecular  volume, the diffusion  constant
$D$ is given by  the  probability of  finding  a free volume equal  to
$v_0$; this leads  to an expression for  $D$, and by extension for the
viscosity $\eta$,

\begin{equation}
\label{eq:5}
\eta\propto 1/D\propto \exp\left(\gamma \frac{v_0}{\overline{v_f}}\right),
\end{equation}
which is similar to the formula first proposed by Doolittle.\cite{R47}

In   the Cohen-Turnbull  formulation,  the  average   free volume  per
molecule is given by $\overline{v_f}=v-v_0$, where $v=1/ \rho$ is the average
total volume per molecule.  The  free-volume concept, in zeroth order,
relies on  a hard-sphere picture in  which thermal activation plays no
role.  For application to real liquids, temperature enters through the
fact  that molecules, or molecular segments  in the  case of polymers,
are  not truly ``hard'' and   that, consequently, the  constant-pressure
volume is temperature-dependent,
\begin{equation}
\label{eq:6}
\overline{v_f(T)}\propto  \alpha_P(T-T_0),
\end{equation}
where $\alpha_P$  is the coefficient of  isobaric expansivity and $T_0$ is
the temperature  at which all free volume  is consumed, i.e., $v=v_0$.
Inserting     the  above   equation   in    the   Doolittle   formula,
Eq.~(\ref{eq:4}),   gives the VFT   expression, Eq.~(\ref{eq:1}).   An
unanswered, but fundamental  question associated with Eq.~(\ref{eq:6})
is why  the free volume should  be consumed at a  nonzero temperature,
$T_0$?   An extended version   of  the free-volume approach has   been
developed by  Cohen   and Grest, in which   the  cages or  ``cells'' are
divided into two groups, liquid-like and solid-like, and concepts from
percolation  theory are included to  describe  the dependence upon the
fraction of liquid-like  cells.  \cite{R48}  (See  also the model  for
molecular diffusivity in  fluids of long rod  molecules by Edwards and
Vilgis.\cite{R49})

The main criticisms  of the free volume models are
(i) that the concept of free volume is ill-defined, which results in a
variety   of  interpretations and   difficulty    in finding a  proper
operational procedure even for simple model systems, and (ii) that the
pressure dependence of  the viscosity (and  $\alpha$-relaxation  times) is  not adequately
reproduced.    This   latter  feature  has  been   emphasized  in many
studies,  \cite{R17,R34,R50} and     it is a
consequence of the observation made above (see  II-3) that the viscous
slowing down of  glass-forming liquids at  1 atm and more generally at
low pressure is primarily controlled by temperature and not by density
or  volume.     Glass  formation  in  supercooled   liquids  does  not
predominantly results from  the drainage of   free volume, but  rather
from thermally activated processes.

\subsection{Mode-coupling approximations}
The   theory   of  glass-forming liquids that      has had the highest
visibility  for  more than    a      decade is  the  mode     coupling
theory. \cite{R51}   It predicts a  dynamic   arrest of the
liquid structural  relaxation  without any  significant  change in the
static properties. All  structural  quantities are  assumed  to behave
smoothly and jamming results from  a nonlinear feedback mechanism that
affects the relaxation  of the   density fluctuations. Formally,   the
theory involves an analysis of a set of nonlinear integro-differential
equations describing  the evolution of  pair  correlation functions of
wave-vector-  and time-dependent  fluctuations that  characterize  the
liquid.    These equations have     the form  of  generalized Langevin
equations, and  they    can be  derived   by  using  the  Zwanzig-Mori
projection-operator formalism. The equation for  the quantity of prime
interest in  the theory, the (normalized)  correlation function of the
density fluctuations,

\begin{equation}
\label{eq:7}
\phi_Q(t)= \frac{<\rho_Q(t)\rho_Q^*(0)>}{<|\rho_Q(0)|^2>},
\end{equation}
where $\rho_Q(t)=\sum_j\exp(i{\bf Q}{\bf r}_j)$
 and ${\bf r}_j$
denotes the position of the $j$th particle, can be written as
\begin{equation}
\label{eq:8}
\frac{d^2}{dt^2}\phi_Q(t)+\Omega^2_Q\phi_Q(t)+\int_0^tdt'm_Q(t-t') \frac{d}{dt'}\phi_Q(t')=0,
\end{equation}
where  $\Omega_Q$ is  a microscopic  frequency obtainable  from the static
structure   factor,  $S(Q)\propto<~|\rho_Q(0)|^2> $,    and  $m_Q(t)$   is the
time-dependent   memory function that   is  formally  related to   the
correlation function of a Q-dependent random force. The above equation
being exact, the crux    of  the mode-coupling approach   consists  in
formulating an approximate expression for $m_Q(t)$.  The mode-coupling
scheme has  been implemented for  liquids both  in  the frame  of  the
kinetic   theory  of fluids \cite{R51}   and  that of  the fluctuating
nonlinear hydrodynamics.   \cite{R52} It   essentially boils down   to
approximating the  memory function  $m_Q(t)$  as the   sum of a   bare
contribution  coming    from   the fast    relaxing variables   and  a
mode-coupling  contribution coming  from the slowly  decaying bilinear
density modes,
\begin{equation}
\label{eq:9}
m_Q(t)=\gamma\delta(t)+\sum_{{\bf Q'}}V_{\bf QQ'}\phi_{\bf Q'}(t)\phi_{|{\bf Q-Q'}|}(t),
\end{equation}
where the vertices $V_{\bf QQ'}$  can be expressed  in terms of the static
structure factor. The self-consistent solution of the resulting set of
nonlinear  equations  predicts a  slowing of   the relaxation  that is
attributed, within a purely homogeneous picture (see  II-2), to a cage
effect and to  the feedback mechanism  above mentioned.  This solution
exhibits a dynamic arrest at a critical point, $T_c$, which represents
a transition from an ergodic to a nonergodic state with no concomitant
singularity in  the thermodynamics and structure   of the system.  The
main achievements of   the  mode-coupling approach are the   predicted
anomalous increase in relaxation time and the appearance of a two-step
relaxation process with  decreasing temperature, as indeed observed in
real  fragile glass-formers  (compare  Fig.   10 to  Fig.   7)  and in
molecular dynamics simulations.  \cite{R39} Early  on, however, it was
realized,  both  from empirical fits   to  experimental data and  from
comparison  to simulation data    on model systems,  that  the dynamic
arrest at  $T_c$  did not describe the   observed glass  transition at
$T_g$ nor  the transition to an ``ideal  glass'' at  a temperature below
$T_g$.   This  is  illustrated in    Fig.  11.  Thus,  the  $T_c$  was
interpreted as a temperature above $T_g$.  The singularity at $T_c$ is
avoided because   of    the breakdown   of  the  simple  mode-coupling
approximation, Eq.  9, and the $T_c$ of what is called the ``idealized''
mode-coupling  theory is taken  as  a crossover below which additional
relaxation mechanisms, such  as  activated processes, presumably  take
over.  Unfortunately, beyond some empirical introduction,
\cite{R51,R52} activated processes are not theoretically described by
mode-coupling approaches, and so the theory of $\alpha$ relaxation has not
been extended  to  temperatures below $T_c$.   To  draw once   again a
parallel with critical phenomena (where  a singularity occurs at $T_c$
in the structure and the thermodynamics  of the system), mode-coupling
approximations, as formulated for  instance by Kawasaki, are known  to
describe  quite well the standard critical  slowing down,  but not the
activated  dynamic scaling such as that  observed in  the random field
Ising  model  (see section   I-1).  This  failure  is related  to  the
underlying nature of the approximation  that corresponds to a one-loop
self-consistent resummation scheme in a perturbative treatment
\cite{R52,R54} (see also below in III-4 the parallel with spin glass models).

Mode-coupling approaches can thus   describe at best the  dynamics  of
moderately supercooled liquids\footnote{It is  possible that  they are
also    applicable  to  the     fast-$\beta$    relaxations  even   below
$T_c$.\cite{R51}}  (see Fig.     11).   Because of the many    detailed
predictions it makes  in this  regime,  the mode-coupling theory   has
stimulated the  use  and the  development of  experimental techniques,
such as neutron and   depolarized   light scattering, and    molecular
dynamics  simulations that are  able to probe   the early stage of the
viscous slowing down;  but, the very  fact that the predicted  dynamic
singularity is not observed makes it difficult  to reach any clear-cut
conclusion about the quantitative adequacy of the theory, and this has
led to  much debate in   recent years. \cite{R55}

\subsection{ Configurational entropy  and (free) energy landscape }
The existence of a crossover temperature in the moderately supercooled
liquid  region   where $\alpha$-relaxation  times  are    of the order  of
$10^{-9}$ sec (hence in the same region as  the $T_c$ predicted by the
mode-coupling   theory)   was      advocated    30  years    ago    by
Goldstein.\cite{R56} Goldstein  argued that  below this crossover flow
is dominated by  potential energy barriers that  are  high compared to
thermal  energies and slow relaxation  occurs as a result of thermally
activated  processes   taking the system   from  one   minimum of  the
potential energy  hypersurface to another.    The idea that  molecular
transport in viscous liquids approaching the glass transition could be
best described by invoking motion of the representative state point of
the system on   the   potential energy   hypersurface had  also   been
suggested by  Gibbs. \cite{R57}  In his  view,  the   slowing down of
relaxations  with decreasing temperature  is  related to a decrease of
the number of available minima  and to  the increasing difficulty  for
the system to find such  minima.  The viscous  slowing down would thus
result from the decrease  of some ``configurational  entropy'' that is a
measure  of the  number   of accessible minima.   These  two concepts,
potential energy hypersurface,   also denoted ``energy  landscape'', and
``configurational entropy'', have   gained a renewed interest in  recent
years, boosted  by  the analogy  with   the  situation encountered  in
several generalized spin   glass models (see  below).  

The Adam-Gibbs approach \cite{R58}
represents a phenomenological attempt to
relate the  $\alpha$-relaxation time of a glass-forming liquid
to  the ``configurational entropy''.   In  the picture,
 $\alpha$-relaxation takes  place by increasingly
cooperative rearrangements of groups  of  molecules.  Any such  group,
called a ``cooperatively   rearranging  region'', is  assumed to   relax
independently  of the    others.  It   is    a  kind   of   long-lived
heterogeneity.  Molecular    motion is  activated   and the  effective
activated free  energy is  equal  to  the  typical energy barrier  per
molecule, which is taken as   independent  of temperature, times   the
number   of molecules that   are  necessary  to  form a  cooperatively
rearranging   region whose   size    permits a  transition   from  one
configuration to  a  new one independently  of  the environment.  This
latter number goes as the  inverse of the configurational entropy  per
molecule,  $S_c(T)/N$, where  $N$ is the total number
of molecules in the sample.  Since $S_c$
 decreases with decreasing temperature, the reasoning leads
to an effective   activation  free energy that  grows  with decreasing
temperature, i.  e., to a super-Arrhenius behavior,
\begin{equation}
\label{eq:10}
\tau_\alpha=\tau_0\exp\left(\frac{C}{TS_c(T)}\right), 
\end{equation}
where $C$ is proportional to $N$ times  the typical energy barrier per
molecule.  If  the    configurational entropy vanishes   at a  nonzero
temperature,   an   assumption    somewhat  analogous   to     that in
Eq.~(\ref{eq:6}) for the  free volume model, but  one that is inherent
for instance in the  Gibbs-di Marzio approximate mean  field treatment
of a lattice  model of linear  polymeric  chains, \cite{R59} then  the
$\alpha$-relaxation times diverge  at  this same nonzero temperature.   In
particular if the configurational entropy is identified as the entropy
difference     between     the    supercooled     liquid    and    the
crystal\footnote{This  phenomenological choice  for   the   entropy of
configuration has been criticized  by Goldstein who showed for several
glass-formers that only half   of the entropy difference  between  the
liquid and the crystal comes  from strictly ``configurational'' sources;
the remainder comes  mostly from changes in  vibrational anharmonicity
or differences  in the number  of molecular  groups able to  engage in
local motions.  \cite{R60}},  the   Adam-Gibbs theory allows   one  to
correlate  the  extrapolated divergence   of the $\alpha$-relaxation times
with  the Kauzmann paradox (see I-4):   the Kauzmann temperature $T_K$
would then signal   a singularity  both in  the  dynamics and  in  the
thermodynamics of  a supercooled liquid\footnote{A recent careful, but
conjectural analysis of dielectric relaxation data suggests that these
data  are  consistent with the  existence  of  a critical  point, both
structural  and dynamical, at the approximate  $T_0$  specified by the
VTF expression in  Eq.~(\ref{eq:1}).\cite{R61}}.   Note  also that  by
using a hyperbolic temperature dependence to fit the experimental data
on the heat  capacity difference between the  liquid and  the crystal,
$\Delta C_P(T)=K/T$,  and using
this  formula to extrapolate the  configurational  entropy down to the
Kauzmann temperature, one converts Eq.~(\ref{eq:10}) to a VFT formula,
\begin{equation}
\label{eq:11}
\tau_\alpha=\tau_0\exp\left(\frac{CT_K}{K(T-T_K)}\right), 
\end{equation}
with  the  VFT  temperature $T_0$ equal    to the Kauzmann temperature
$T_K$.      When      comparing    to    experimental     data,    the
configurational-entropy based  expressions  provide a good description
at  least over  a  restricted temperature   range, but  the  resulting
estimates  for   the   critical   number  of molecules    composing  a
cooperatively   rearranging region is  often  found to be unphysically
small (only a few molecules at $T_g$).
\cite{R1}

Building  upon the early suggestion  made by Goldstein, \cite{R62} and
others proposed that the apparent  passage with decreasing temperature
from flow dynamics described by  a mode-coupling approach to activated
dynamics such as pictured  by  the configurational-entropy theory   of
Adam  and Gibbs  could be rationalized   by considering the physics of
exploration  of the  energy   landscape:  see  Fig.  12.  The   energy
landscape is the potential energy  in configurational space. It can be
envisaged as an incredibly complex, multi-dimensional ($3N$ dimensions
for a   system of  $N$  particles)  set  of hills,   valleys,  basins,
saddle-points, and passage-ways around  the hills.  At constant volume
and constant number of  particles,  this landscape is  independent  of
temperature.   However, the fraction of   space that is  statistically
accessible to the  representative state point  of the system decreases
with decreasing  temperature, and  the  system becomes constrained  to
deeper  and deeper wells. (Recall  that   below the melting point  the
deepest   energy minima  corresponding   to  the crystalline   part of
configurational  space must be excluded  when studying the supercooled
liquid.)  At low enough  temperature, when the representative point of
the supercooled   liquid is  mostly  found in   fairly deep and narrow
wells, it seems reasonable to  define a ``configurational entropy'' that
is proportional  to  the logarithm of  the number  of minima  that are
accessible   at  a   given   temperature.  The  liquid  configurations
corresponding to  these accessible minima   have been called ``inherent
structures''       and Stillinger  and      coworkers   have devised  a
gradient-descent mapping procedure to find the inherent structures and
study   their   properties   in    computer  simulations.   \cite{R63}
Interestingly, Stillinger    has  also  shown,  with    fairly general
arguments, that   if one  is    to use  the  above defined   notion of
configurational entropy,   an ``ideal glass   transition'' of  the  type
commonly  associated   with    the Kauzmann  paradox,    i.   e.,  one
characterized by the vanishing   of the configurational entropy at   a
nonzero temperature, cannot occur  for  systems of limited   molecular
weight  and short-range   interactions. \cite{R64} 

It may be  more  fruitful to investigate  in  place  of the  potential
energy landscape a free-energy landscape. Such a landscape can only be
defined if  one is  able to  construct a free-energy  functional  by a
suitable coarse-graining procedure, as can be done for instance in the
case  of mean-field  spin glass  models   (see below).  A  free-energy
landscape is temperature-dependent, and  it is important to note  that
the ``configurational  entropy'',  also  called  ``complexity'',\cite{R65}
that  one  can define  from the   logarithm of  the number  of
accessible free-energy  minima  differs  from   the   ``configurational
entropy'' computed from the potential energy landscape.  In particular,
the   behavior of the complexity is   not restricted by the Stillinger
arguments given above. 

The ``landscape paradigm''  is  very  appealing  in  rationalizing  many
observations    on  liquids and    glasses and,   more   generally, in
establishing a framework   to describe qualitatively  slow dynamics in
complex systems that span a wide range of scientific fields.\cite{R66}
It has been used to motivate, in addition to the Adam-Gibbs theory and
other   phenomenological approaches  like  the  soft-potential  model,
\cite{R67} simple   stochastic  models  of    transport based   on  master
equations.  \cite{R68} Nevertheless, it has not so far offered a
way for elucidating the  {\it physical} mechanism that is  responsible
for  the  distinctive    features of the   viscous   slowing   down of
supercooled liquids.  

\subsection{Analogy with generalized spin glass models}
If  one  takes   seriously  the  observation   that  the  extrapolated
temperature dependence   of  both the  viscosity  (and $\alpha$-relaxation
times) and  the ``configurational'' entropy (taken  as the difference of
entropy    between the liquid and   the   crystal) become divergent or
singular at essentially the same temperature $T_0\simeq T_K$(see Fig.  9),
one is naturally led to postulate the existence at this temperature of
an underlying thermodynamic  transition,  usually referred  to  as the
``ideal  glass  transition''.   Looking    for  analogies  with    phase
transitions in  spin glasses is  then appealing.  However, the kind of
dynamic  activated scaling   that  would be  required to  describe the
slowing   down  of relaxations    when   approaching the   ideal glass
transition  (see I-1)  is not  found in  the  most studied Ising  spin
glasses.  \cite{R69} Kirkpatrick, Thirumalai,  and Wolynes argued that
generalized spin  glass models, such as Potts  glasses and random {\it
p}-spin  systems,  would   be  better candidates.   \cite{R70,R71} The
random  {\it p}-spin model, for instance,  is defined by the following
hamiltonian:
\begin{equation}
\label{eq:12}
H=\sum_{i_1<i_2<\ldots<i_p}J_{i_1i_2\ldots i_p}\sigma_{i_1}\sigma_{i_2}\ldots\sigma_{i_p},
\end{equation}
where the    $\sigma_{i}$'s   are  Ising  variables   and    the couplings
$J_{i_1i_2\ldots  i_p}$'s are quenched  independent  random variables that
can take positive and negative values according to a given probability
distribution.    The behavior of  these  systems,  {\it at least  when
solved in  the mean-field limit where   the interactions between spins
have  infinite range},  bears many  similarities  with the theoretical
description    of  glass-forming  liquids    outlined above.   Indeed,
mean-field Potts glasses (with a number of states strictly larger than
4) and mean-field  p-spin models (with $p\geq  3$) have essentially  the
following characteristics:
\begin{enumerate}
\item
 At
high temperature, the system  is in a fully  disordered (paramagnetic)
state. At   a  temperature   $T_D$,   there  appears   an
exponentially large number of metastable ``glassy'' states whose overall
contribution to the  partition  function  is  equal  to that of    the
paramagnetic minimum. The free energy and all other static equilibrium
quantities are fully regular at   $T_D$, but  the dynamics have a
singularity of the exact same type as that  found in the mode-coupling
theory of     liquids.  At   $T_D$,  the system is trapped in
one of the metastable free-energy  minima, and ergocity is broken.

\item   Below   $T_D$ ,  a peculiar   situation  occurs.  The
partition  function  has  contributions from,   both, the paramagnetic
state  and  the exponentially  large  number of ``glassy'' (free-energy)
minima, the logarithm of which  defines the configurational entropy or
complexity.  This latter  decreases as the temperature is lowered.
\item
 At   a nonzero temperature  $T_s<T_D$,
 the configurational entropy  vanishes. The system undergoes a
{\it bona  fide }thermodynamic transition to a  spin-glass phase.  The
transition has been termed ``random first order'' 
\cite{R70,R71} because it is  second-order  in the usual thermodynamic sense
(with, e. g., no  latent heat), but  shows a discontinuous jump in the
order  parameter.  (Technically,   within the  replica  formalism,  it
corresponds     to   a one-step  replica   symmetry    breaking with a
discontinuous  jump  of  the Edwards-Anderson   order
parameter.\cite{R69,R70,R71})
\end{enumerate}

These   mean-field systems  are  the  simplest, analytically tractable
models found so  far  that  display a high-temperature   mode-coupling
dynamic  singularity,   a   nontrivial  free-energy   landscape, and a
low-temperature  ideal  (spin)  glass  transition    with an  ``entropy
crisis''\footnote{They  are also  aging phenomena, as  discussed in Ref
\cite{R5}.}.   Analyzing   them   sheds light   on   the  mode-coupling
approximation, whose  validity for fluid  systems is otherwise hard to
assess.    The  mode-coupling  approximation   becomes  exact  in  the
mean-field  limit,   because the  barriers   separating the metastable
minima diverge  (in the thermodynamic limit)  at and below  $T_D$ as a
result of the assumed infinite  range of the interactions. One expects
that in  a finite-range model,  provided the  same type of free-energy
landscape  is  still encountered, barriers are    large but finite and
ergodicity is  restored by thermally activated processes. Accordingly,
the  dynamic transition is  smeared out, and  the activated relaxation
mechanisms that take over must be  described in a nonperturbative way,
as suggested for instance by Kirkpatrick, Thirumalai, and Wolynes
\cite{R71}   in  their  dynamic   scaling approach  based  on entropic
droplets.  

An   advantage of  an   analogy  between   glass-forming liquids  and
generalized spin  glasses\footnote{See also the frustrated percolation
model\cite{R72}.} is that the powerful tools  that have been developed
in the theory of spin-glass models to characterize the order parameter
and the properties associated with the  existence of a large number of
metastable glassy states, among which the replica formalism,
\cite{R69} can be used {\it  mutatis mutandis }to{\it }study liquids and
glasses.   \cite{R73} However, to  make the analogy really successful,
one must still find a short-range model (even more convincing would be
a model  without  quenched disorder) that  actually displays activated
dynamic scaling and a random  first-order transition and make progress
in describing the slow relaxation.

\subsection{Intrinsic frustration  without randomness }
Spin glasses, and  related systems like orientational  glasses, vortex
glasses  and  vulcanized  matter,  \cite{R74}   owe their  fascinating
behavior   to  two  main ingredients:  {\it    randomness}, namely the
presence    of an   externally  imposed quenched   disorder, and  {\it
frustration}, which expresses   the impossibility   of  simultaneously
minimizing  all the interaction terms  in  the energy  function of the
system.  Liquids and glasses (sometimes called "structural glasses" to
stress the difference with spin  glasses) have no quenched randomness,
but frustration has been  suggested as  a  key feature to  explain the
phenomena  associated with glass formation. \cite{R75,R76,R77,R78,R79}
Frustration in  this context is attributed to  a competition between a
short-range tendency  for the extension of  a  locally preferred order
and global constraints that preclude  the periodic tiling of the whole
space with the  local structure. 

The best  studied example  of  such an intrinsic frustration  concerns
single-component  systems  of   spherical   particles interacting with
simple  pair  potentials.   What is   usually  called  "geometric"  or
"topological"  frustration can be  more easily understood by comparing
the situations encountered in  $2$ and $3$ dimensions  \cite{R76} (see
Fig.  13).  In  $2$  dimensions,  the  arrangement  of disks   that is
locally preferred, in  the  sense that  it maximizes the  density  and
minimizes the energy, is  a hexagon of  6 disks around a central  one,
and  this hexagonal structure  can be  extended to the  whole space to
form a triangular  lattice.  In $3$ dimensions, as  was shown long ago
by Frank, \cite{R80} the locally  preferred cluster  of spheres is  an
icosahedron; however, the  $5$-fold rotational symmetry characteristic
of icosahedral order is incompatible  with translational symmetry, and
formation of a periodic  icosahedral crystal is forbidden.   Geometric
or topological frustration is thus  absent in the $2$-dimensional case
but present in  the $3$-dimensional case.   A consequence of this, for
instance,  is that      crystallization is  continuous,    or   weakly
first-order, in  $2$   dimensions  (with some  subtleties   related to
ordering  in   $2$  dimensions  \cite{R81}) whereas  it    is strongly
first-order in $3$  dimensions and accompanied  by the breaking of the
local  icosahedral  structure  to   make the    face-centered-cubic or
hexagonal-close-packed order that  allows to tile space  periodically.
(In contrast, aligned cubes  in 3 dimensions  have no frustration  and
undergo a  continuous   freezing transition to  a  crystalline  state.
\cite{R82}).  The   geometric  frustration   that  affects spheres  in
$3$-dimensional Euclidean space can be relieved in curved space with a
specially  tuned  curvature;   the  creation  of  topological  defects
(disclination lines)  can then be viewed  as the result of forcing the
ideal icosahedral ordering into "flat" space.   This picture of sphere
packing disrupted by frustration  has been further developed in models
for simple atomic systems and metallic glasses, \cite{R75,R76} and the
slowing down of relaxations  has  been tentatively attributed  to  the
topological constraints that   hinder  the kinetics of  the  entangled
defect    lines; \cite{R76} however,  the  treatment  remains only
qualitative and incomplete.

A significant  difficulty in   applying  the concept of  geometric  or
topological frustration  to supercooled liquids   is that real fragile
glass-formers  are in  general   either mixtures  or  single-component
systems  of nonspherical  molecules with  a variety of  shapes, all of
which obscures the  detailed   mechanisms  and constraints that    are
responsible  for  the  frustration.  Attempts  have been   made to get
around this problem by  proposing a more coarse-grained description of
frustration\footnote{In addition, several ``toy models''possessing
frustration, but no quenched disorder have been studied by computer
simulation: see for instance Ref.\cite{R83}.}. \cite{R77,R78,R79}

In Stillinger's "tear and  repair" mechanism for relaxation and  shear
flow   \cite{R78}    and   in     the   more      recently  introduced
"frustration-limited     domain    theory",\cite{R79}  frustration  is
described  as the source  of a  strain  free  energy that  opposes the
spatial extension  of   the  locally  preferred structure   and  grows
super-extensively with system size.  It  results in the breaking up of
the  liquid  into   domains, whose size   and  growth  with decreasing
temperature are limited by frustration, the weaker the frustration the
larger the domains.  The super-Arrhenius temperature dependence of the
viscosity and $\alpha$-relaxation times   and the heterogeneous nature  of
the   dynamics  are  attributed     to  these  domains   (see     also
Ref.\cite{R77}).  Progress has been  made along these lines, by making
use of  a  scaling approach based on  the  concept of avoided critical
behavior\footnote{This   approach differs from   both those based upon
spin-glass  analogies  and  those  in  which   the slow kinetics   are
attributed to frustration-induced entangled defect lines in that these
others scale about a  low-temperature characteristic point  signifying
ultimate  slowing  down, whereas  in  the  frustration-limited  domain
theory    the scaling is    carried   out  about  a   high-temperature
characteristic  point  signifying the initiation  of  anomalously slow
dynamics.  For the  same reason, it also  differs from the  domain (or
cluster) picture that  has been proposed on  the  basis of  an analogy
between a supercooled liquid  approaching the glass transition  and a
mean-field  model   with   purely  repulsive interactions    near  its
spinodal.\cite{R85}}.    \cite{R79,R84}   However, the  putative order
variable characterizing the locally  preferred structure of the liquid
has  not yet been   properly identified, and,   as in the  case of the
generalized spin-glass  models  discussed above,  one must  still give
convincing  evidence that  the $3$-dimensional  statistical-mechanical
frustrated models   that have been  suggested  as  minimal theoretical
descriptions do show the expected activated dynamics.

\section{CONCLUSION}
The viscous  slowing down of supercooled liquids  that leads  to glass
formation  can be  considered as  a  classical  and thoroughly studied
example of a "jamming process".  In this review,  we have stressed the
distinctive  features     characterizing  the   phenomenon:    strong,
super-Arrhenius temperature  dependence  of  the  viscosity  and   the
$\alpha$-relaxation  times,  nonexponential and heterogeneous character of
the $\alpha$  relaxation,  absence  of  marked changes   in the structural
(static) quantities, rapid decrease of the  liquid entropy relative to
that of the crystal, appearance of a sequence of steps (or regimes) in
the relaxation   functions.    These  features are   common   to  most
glass-forming liquids (with the exception of  systems forming $2$- and
$3$-dimensional  networks of  strong  intermolecular  bonds).  We have
also discussed the main theoretical approaches that have been proposed
to describe the origin and the nature of the  viscous slowing down and
of the glass transition. We have emphasized the concepts, such as free
volume,    dynamic      freezing  and  mode-coupling   approximations,
configurational entropy and  (free) energy landscape, and frustration,
that could be useful in other areas of physics where jamming processes
are encountered.

\begin{figure}

\caption[99]{
Super-Arrhenius   $T$-dependence  of     the     viscosity $\eta$    and
$\alpha$-relaxation times $\tau_\alpha  $ in  glass-forming  liquids. { \bf  a)}
Logarithm   (base 10)  of  $\eta$  and $\tau_\alpha $   versus reduced inverse
temperature $T_g/T$ for several liquids. For $GeO_2$, a system forming
a  network of strong intermolecular  bonds,  the  variation is  almost
linear, whereas    the  other  liquids   (glycerol,   m-toluidine, and
ortho-terphenyl) are characterized, below some temperature $T^*$, by a
strong  departure from linear    dependence:  the behavior  is  nearly
Arrhenius in the former case and super-Arrhenius  in the latter. (Data
taken   from references      cited  in     Ref.\cite{R8}  and     from
C. Alba-Simionesco, private  communication.)   { \bf b)  }   Effective
activation free energy  $E(T)$, obtained from data  shown in {\bf a)},
as a function of inverse temperature. Both $E(T)$  and $T$ are divided
by the crossover temperature $T^*$ shown in { \bf a) }.}

\end{figure}

\begin{figure}
\caption[99]{
Imaginary  part  of  the dielectric  susceptibility $\chi''$ of liquid
m-toluidine            versus           $ log_{10}(\omega)$ for several
temperatures close to $T_g$ ($T_g=183.5 K$). The inset
shows       that  the $\alpha$ peak  is  broader than a Debye spectrum
that     would correspond  to    a  purely   exponential relaxation in
time.     (Data   from     C.    Tschirwitz,    E. R{\"o}ssler,   and
C.   Alba-Simionesco,  private  communication.)}
\end{figure}

\begin{figure}
\caption[99]{Static
structure  factor   $S(Q)$  of  liquid  (deuterated) ortho-terphenyl  at
several temperatures from just  below melting ($T_m=329 K$) to
just above the glass transition ($T_g=243 K$). The inset shows
the weak  variation with  temperature of  $S(Q)$ for three  values  of $Q$
indicated by the symbols above  the $Q$-axis. (From Ref.\cite{R13}.)}
\end{figure}
\begin{figure}
\caption[99]{
Kauzmann's representation of the ``entropy paradox'': entropy difference
$\Delta S$ between the liquid and the crystal (normalized by its value $\Delta
S_m$ at the  melting  point) versus  reduced temperature  $T/T_m$. The
break in the slopes of the full lines signals  the glass transition at
$T_g$.    The dashed lines indicate    an extrapolation of the entropy
difference curves below $T_g$.  Except for the strong, network-forming
liquid $B_2O_3$,   the extrapolated entropy   difference vanishes at a
nonzero temperature $T_K$.  (Data from Refs.\cite{R16},\cite{R18}, and
from H.  Fujimori and C.  Alba-Simionesco, private communication.)}
\end{figure}
\begin{figure}
\caption[99]{Time  dependence  of the (normalized)
dynamic       structure     factor       $S(Q,t)/S(Q)$      of    liquid
$Ca_{0.4}K_{0.6}(NO_3)_{1.4}$  (for
$Q\simeq 1.9{\AA}^{-1} $) at various
temperatures.  The  continuous   lines    are   obtained  by   Fourier
transforming neutron  time-of-flight data  and  the symbols  represent
neutron spin-echo results. When $T$ decreases, two  steps separated by a
plateau    appear  in the  relaxation.     (From Ref.\cite{R19}.)}
\end{figure}
\begin{figure}
\caption[99]{
Scaling plot  for the     imaginary  part $\chi''$ of  the    dielectric
susceptibility  of  several  glass-forming  liquids  (glycerol, salol,
propylene     glycol,      dibutyl-phtalate,     $\alpha$-phenyl-o-cresol,
ortho-terphenyl,  ortho-phenylphenol). Experimental   data similar  to
those shown in Fig. 2, but for 13 decades  of frequency, are collapsed
for    all temperatures  and    all  liquids  onto  the   master-curve
$w^{-1}log_{10}(\chi''\omega_\alpha/  \Delta\chi\omega)$ vs $w^{-1}(1+w^{-1})log_{10}(\omega/
\omega_\alpha)$; $\omega_\alpha$ is the $\alpha$-peak position, $w$ is a shape factor that
characterizes  the deviation from Debye  behavior,  and $\Delta\chi$ is  the
static susceptibility. (From Ref.\cite{R11}.) }
\end{figure}
\begin{figure}
\caption[99]{``Decoupling'' between rotational and
translational  time scales: logarithm    (base 10) of the   rotational
($D_r$)  and   translational  ($D_t$)  diffusion     coefficients  for
ortho-terphenyl  as a  function of  temperature  and viscosity.  $D_r$
(diamonds) follows the   viscosity at all temperatures whereas   $D_t$
(squares, triangles, and dots) departs from viscosity (and from $D_r$)
below $T\simeq 290 K$.  (From Ref.\cite{R28}).}
\end{figure}
\begin{figure}
\caption[99]{
Relative influence of temperature  ($T$)  and density  ( $\rho$) on  the
viscous  slowing    down of  liquid   triphenyl  phosphite.   { \bf a) }
$log_{10}(\eta)$  versus $\rho$  at   constant $T$ for several  isotherms.
Also shown    are the isobaric   data  at $P=1atm$  (under atmospheric
pressure,  the   glass  transition takes  place at    $T_g\simeq 200K$ and
$\rho_g\simeq1.275g/cm^3$).  {\bf b)} $log_{10}(\eta)$  versus $T$ at constant
$\rho$ for several isochores. Note that the  change of viscosity is much
smaller with density (at constant $T$) than it is with temperature (at
constant $\rho$) for the range of temperature and density characteristic
of  the liquid and supercooled  liquid phases at atmospheric pressure.
(From Ref.\cite{R34}.) }
\end{figure}
\begin{figure}
\caption[99]{Illustration of the various
choices  of   characteristic  temperature for  describing  the viscous
slowing down  of  liquid ortho-terphenyl.  $log_{10}(\eta)$   is plotted
versus inverse temperature.  ``Extrapolation'' temperatures:  $T_0(VFT$,
see Eq.  1)$=200-202  K$ \cite{R18,R35}, $T_K$(Kauzmann, see I-4)=$204
K$  \cite{R18}.   ``Crossover''   temperatures: $T^*$(see   Fig.   1 and
III-5)=$350 K$\cite{R8}, $T_c$(MCT, see III-2)=$276-290 K$
\cite{R38}, $T_A=455 K$ \cite{R36}, $T_B=290 K$ \cite{R37}, $T_X=289
K$\cite{R37}.  Also shown    are the experimentally measured   boiling
($T_b=610  K$), melting ($T_m=311 K$),  and glass transition ($T_g=246
K$) temperatures.  (Viscosity data from Ref.
\cite{R38}.) }
\end{figure}

\begin{figure}
\caption[99]{Mode-coupling scenario of kinetic freezing and appearance
of   a  $2$-step  relaxation:  time dependence   of   the (normalized)
density-density   correlation function for  a  schematic mode-coupling
model.  The curves  from $A$ to $G$  correspond to the approach toward
the dynamic  singularity  from  the ergodic  state.  The other  curves
correspond to the nonergodic state. (From Ref.\cite{R51}.)}
\end{figure}
\begin{figure}
\caption[99]{Breakdown of
the  ``idealized''     mode-coupling theory   illustrated   on  $log_{10}(\eta)$ 
 versus $1/T$  for liquid ortho-terphenyl
(see caption to Fig. 9). The  continuous line is the mode-coupling fit
to the experimental data. The predictions break  down below a point at
which  the viscosity  is  of the   order  of  10  Poise;   the dynamic
singularity is not observed,  and  $T_c$ is interpreted as   a
crossover    temperature.}
\end{figure}
 \begin{figure}
\caption[99]{Illustration of  the  putative relation
between the rapid increase of $\alpha$-relaxation time ({\bf a}), the decrease
of  the entropy difference  between the liquid  and  the crystal ({\bf
b}), and the characteristic energy level  on the (schematic) potential
energy landscape   ({\bf c})  for  a  fragile glass-former  at various
temperatures. The   ideal   glass level  corresponds   to the Kauzmann
temperature  (see I-4) and     to an extrapolated  divergence   of the
relaxation  time;   Goldstein's   crossover (see   III-3)   takes place
somewhat  below point $2$.  (From Ref.\cite{R17}.)}
 \end{figure}
 \begin{figure}
\caption[99]{Illustration  of    geometric
frustration for spherical particles. {\bf a)} Packing in 2 dimensions:
equilateral triangles  are preferred locally  and  combine  to form  a
hexagonal local cluster that can tile space to generate a close-packed
triangular  lattice. {\bf b)}  Packing in $3$ dimensions: tetrahedra are
preferred  locally  and combine (with slight   distortions)  to form a
regular icosahedral cluster; however,  the $5$-fold symmetry axes of the
icosahedron preclude a simple icosahedral space-filling lattice. (From
Ref. \cite{R76}.) }
 \end{figure}
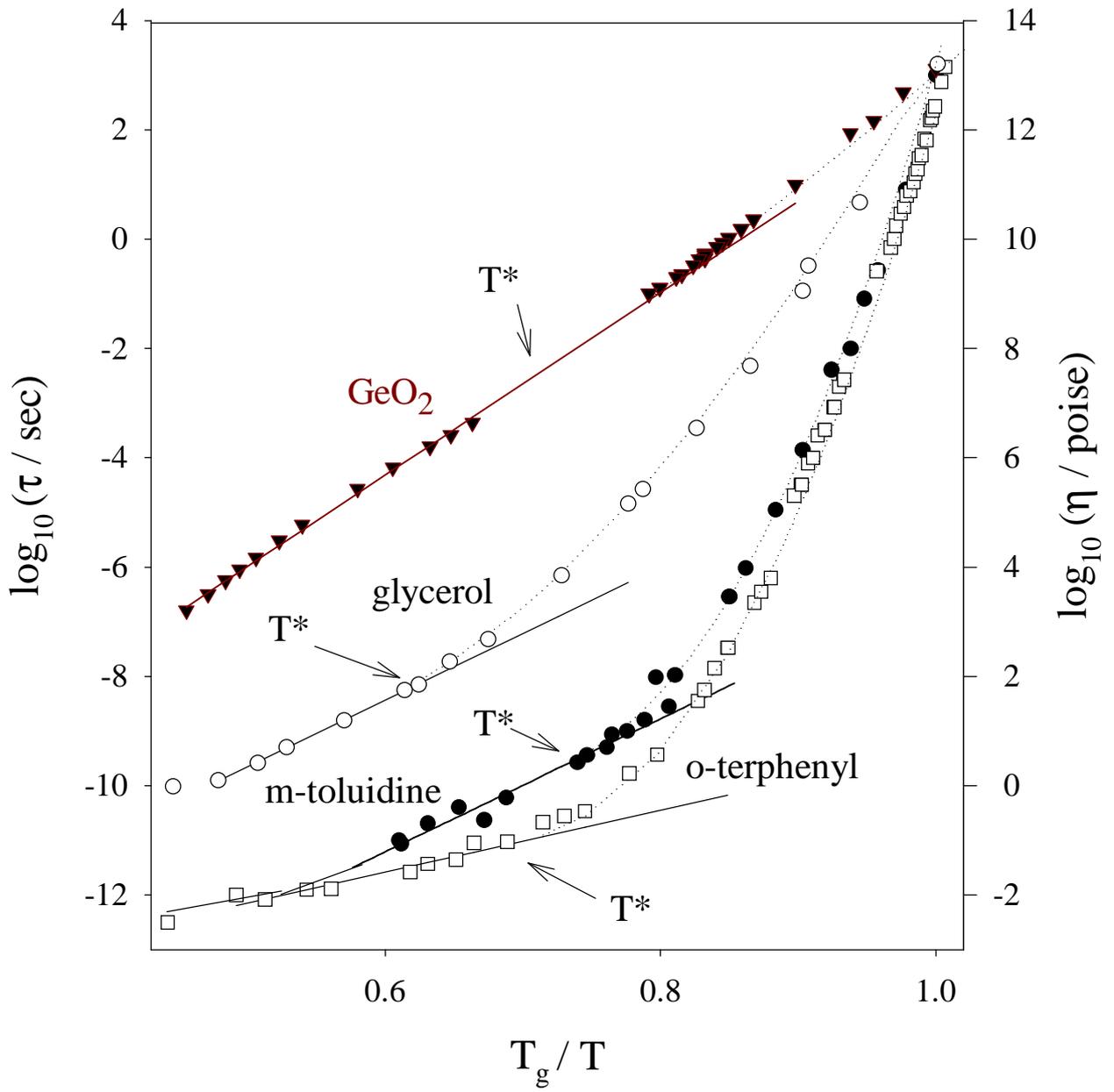
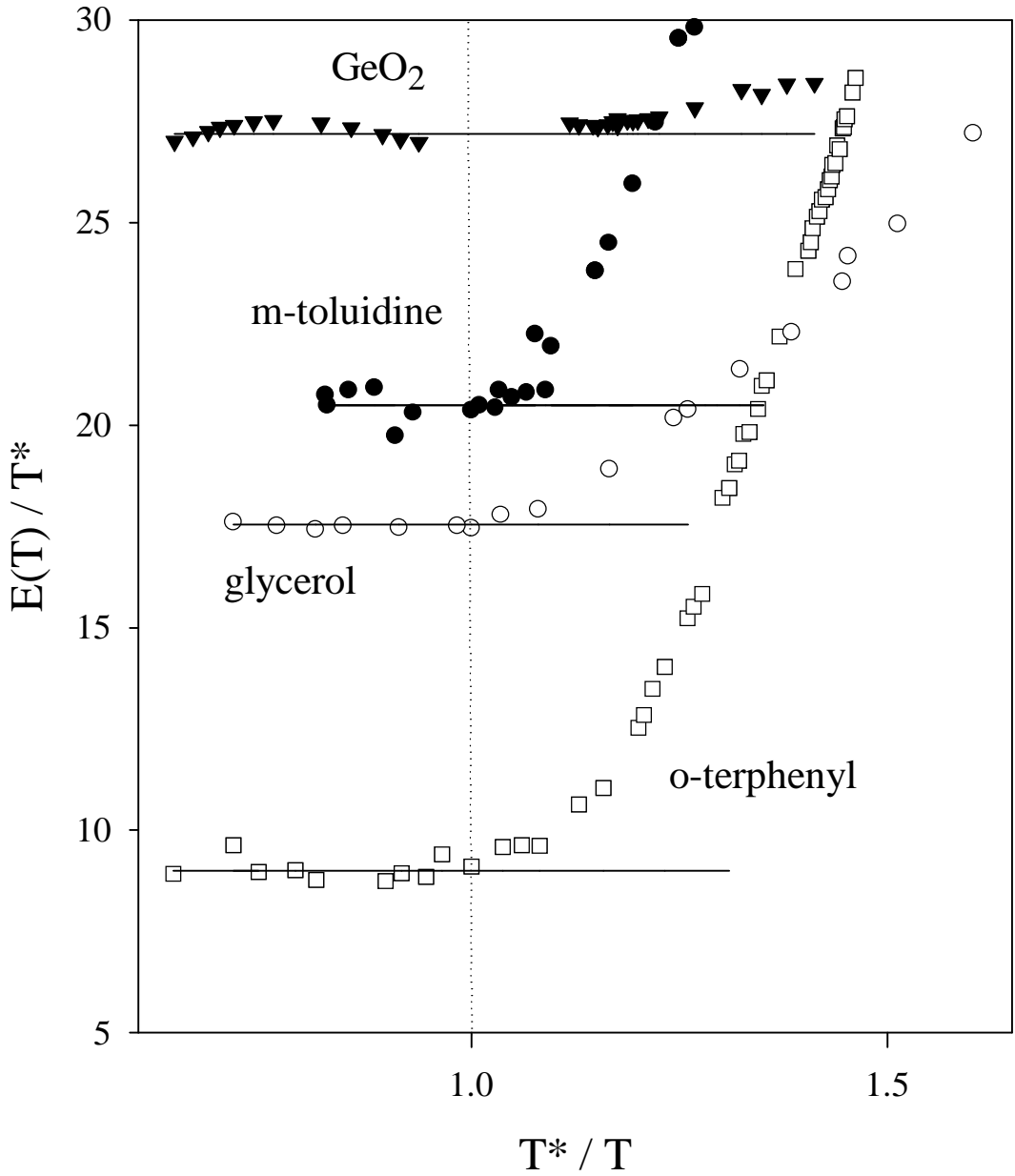
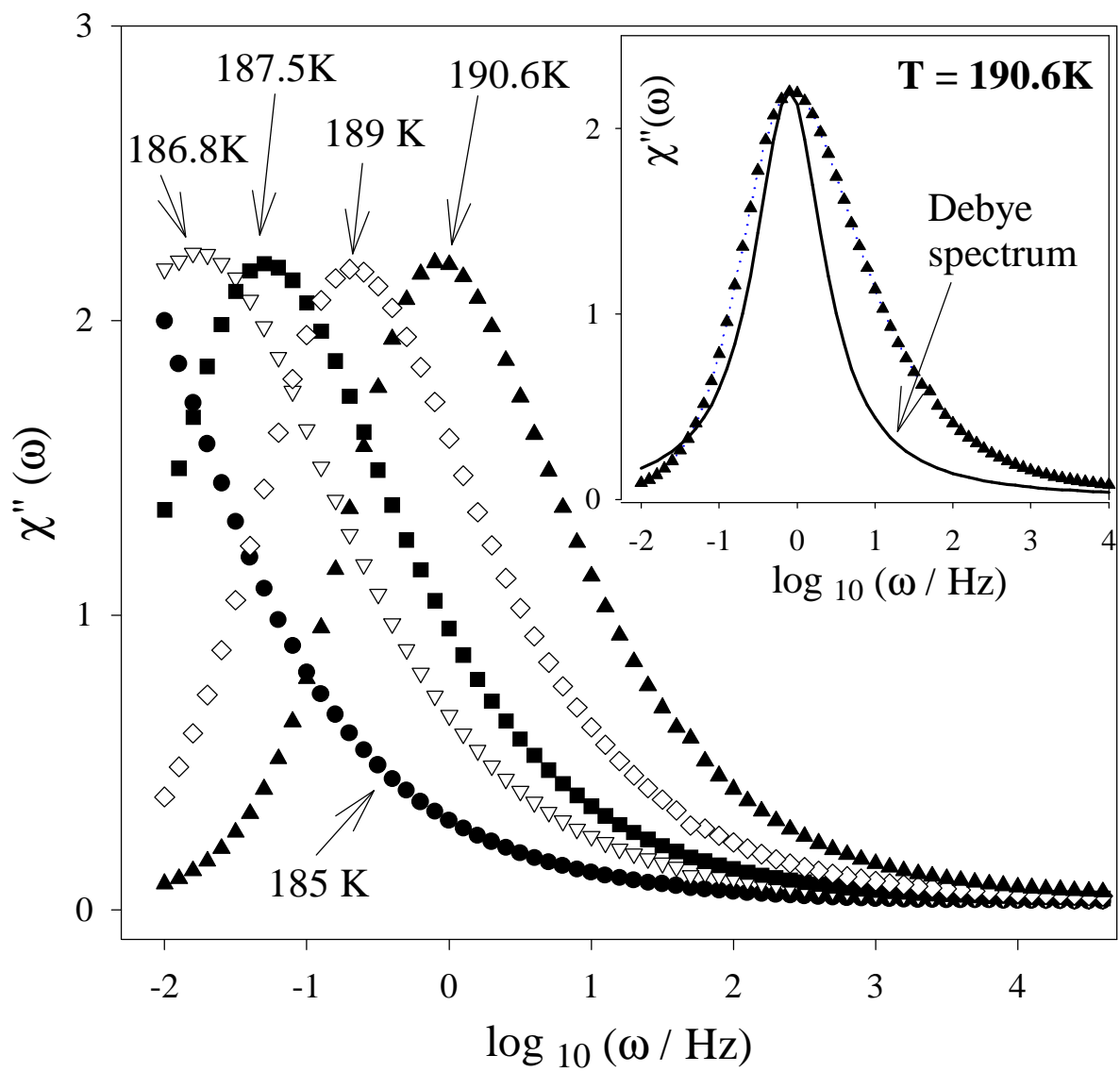
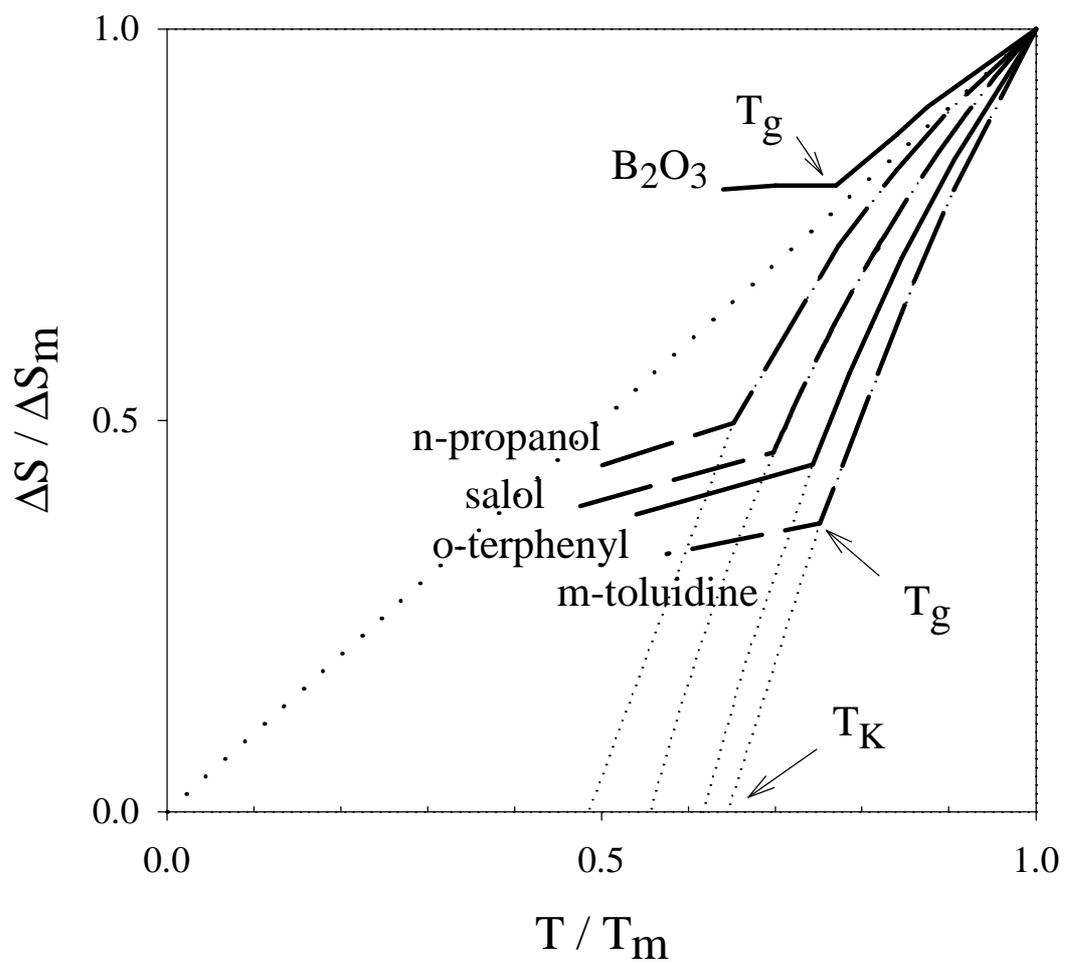
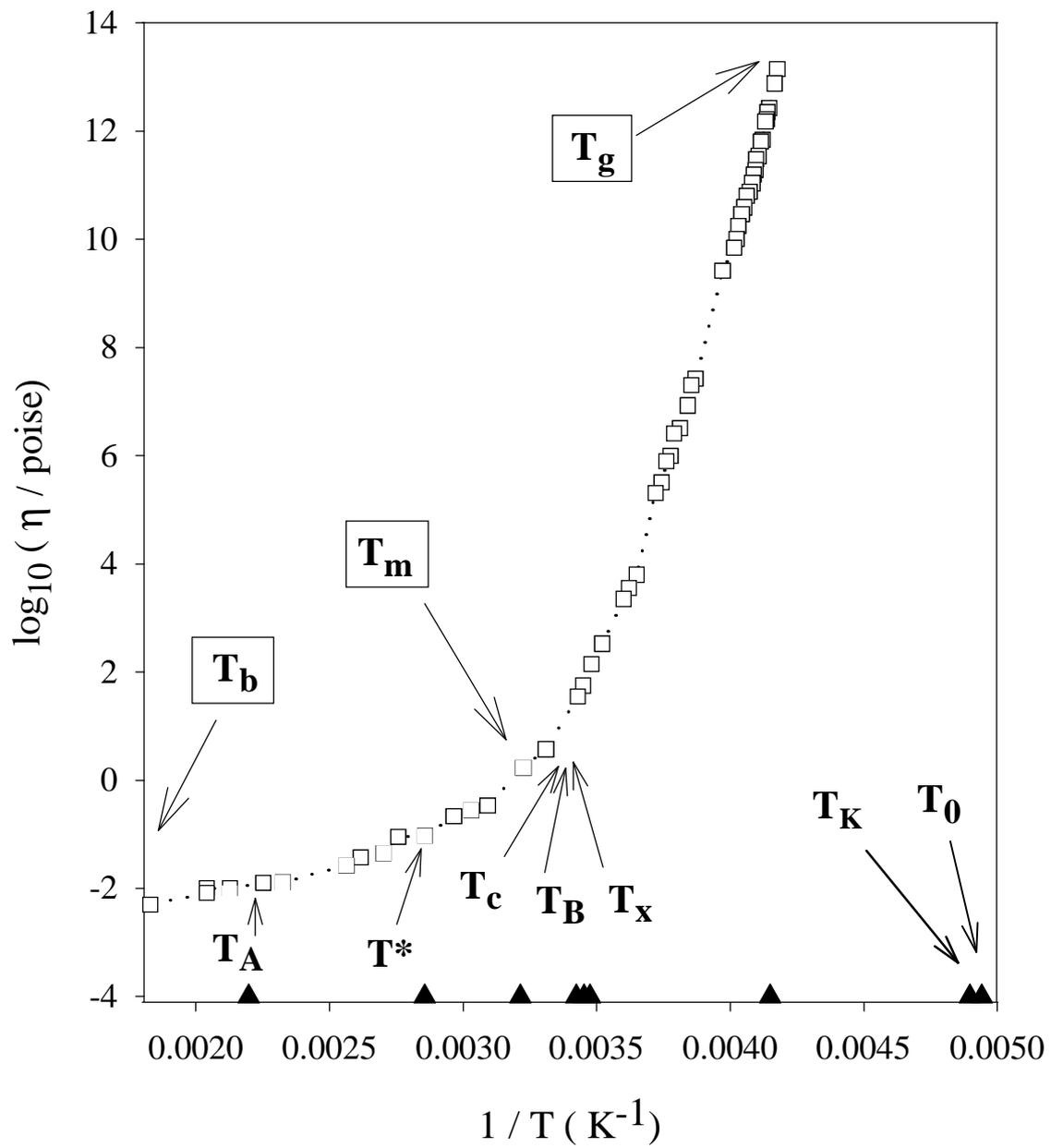
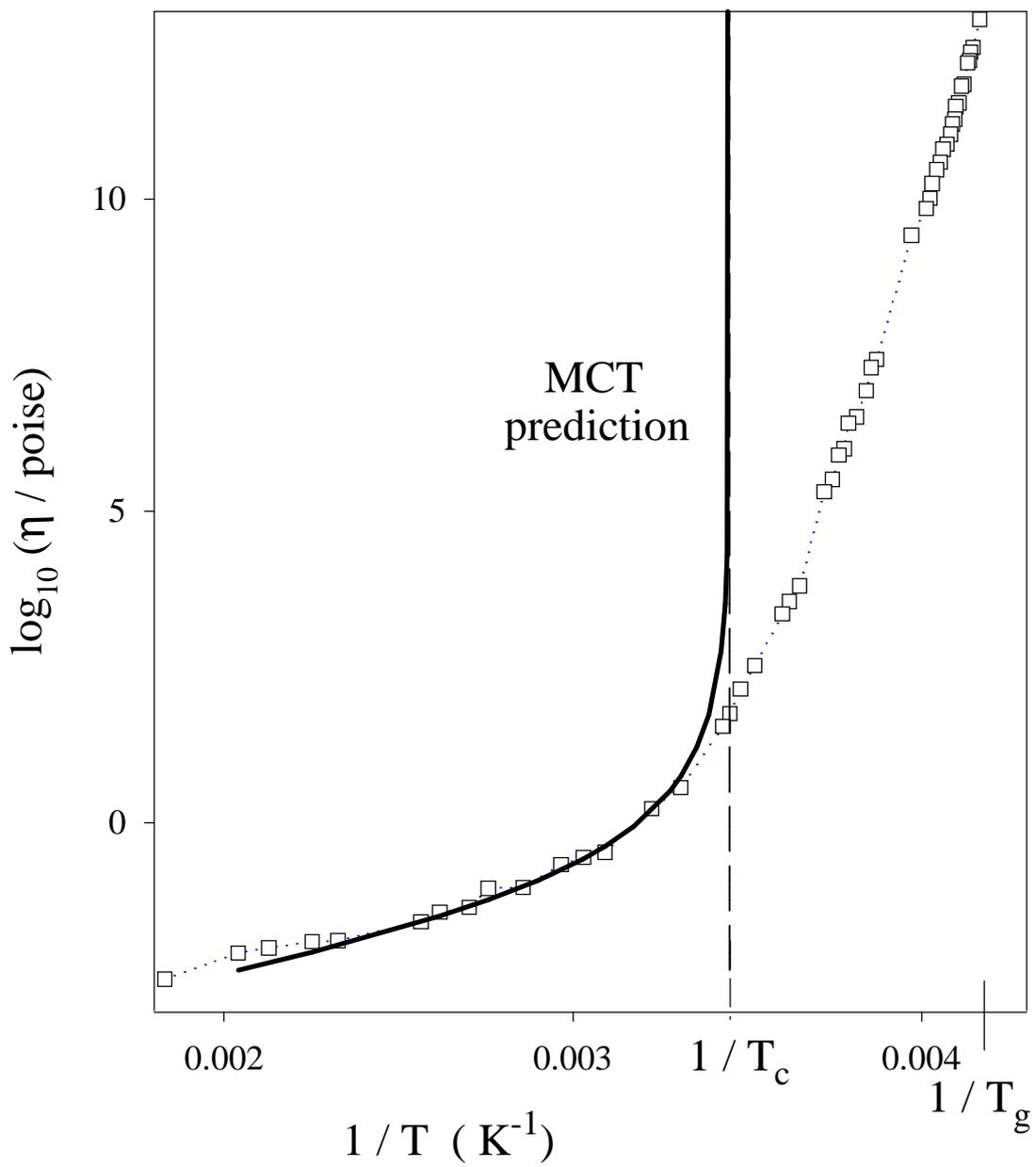
\end{document}